\documentclass[11pt, preprint]{aastex}

\usepackage{epsf,epsfig}
\begin{document}
\newcommand{\dd}{deg$^{2}$}

\title {The XMM-LSS survey. Survey design and first results\thanks{paper based on observations obtained with the
XMM,CFH, ESO (Prg: P70. A-0283; .A-0733), VLA, CTIO and Las
Campanas observatories}}

\author{Marguerite PIERRE, Ivan VALTCHANOV\altaffilmark{1},}
\affil{ CEA/DSM/DAPNIA, Service d'Astrophysique, Saclay,\\
F-91191 Gif sur Yvette\\
       {\it mpierre@cea.fr}}

\altaffiltext{1}{currently at Imperial College, London}

\author{Bruno ALTIERI }
\affil{XMM Science Operations Centre, Villafranca, Spain \\
    {\it }}

\author{Stefano ANDREON }
\affil{Osservatorio Astronomico di Brera, Milano, Italy}

\author{Micol BOLZONELLA \altaffilmark{6}}
\affil{IASF, Milano, Italy} \altaffiltext{6}{currently at
Osservatorio Astronomico di Bologna, Bologna, Italy}

\author{Malcolm BREMER}
\affil{Department of Physics, University of Bristol, UK}

\author{Ludovic DISSEAU}
\affil{European Southern Observatory, Santiago, Chile}

\author{Sergio DOS SANTOS}
\affil{CEA/DSM/DAPNIA, Service d'Astrophysique, Saclay, France}

\author{Poshak GANDHI}
\affil{European Southern Observatory, Santiago, Chile}

\author{Christophe JEAN}
\affil{Institut d'Astrophysique et de G\'eophysique, Universit\'e
de Li\`ege, Belgium }

\author{Florian PACAUD}
\affil{CEA/DSM/DAPNIA, Service d'Astrophysique, Saclay, France  }

\author{Andrew READ\altaffilmark{2}}
\affil{School of Physics and Astronomy, University of Birmingham,
UK} \altaffiltext{2}{currently at University of Leicester}

\author{Alexandre REFREGIER}
\affil{CEA/DSM/DAPNIA, Service d'Astrophysique, Saclay, France}

\author{Jon WILLIS\altaffilmark{3} \altaffilmark{4}}
\affil{European Southern Observatory, Santiago, Chile}
\altaffiltext{3}{ Pontificia Universidad Cat\'olica, Santiago,
Chile} \altaffiltext{4}{currently at University of Victoria,
Canada}
\author{Christophe ADAMI }
\affil{Laboratoire d'Astrophysique, Marseille, France }

\author{Danielle ALLOIN}
\affil{European Southern Observatory, Santiago, Chile}

\author{Mark BIRKINSHAW}
\affil{Department of Physics, University of Bristol, UK}

\author{Lucio CHIAPPETTI}
\affil{ IASF, Milano, Italy}

\author{Aaron COHEN}
\affil{Naval Research laboratory, Washington, US}

\author{Alain DETAL}
\affil{Institut d'Astrophysique et de G\'eophysique, Universit\'e
de Li\`ege, Belgium }

\author{Pierre-Alain DUC}
\affil{CEA/DSM/DAPNIA, Service d'Astrophysique, Saclay, France}

\author{Eric GOSSET}
\affil{Institut d'Astrophysique et de G\'eophysique, Universit\'e
de Li\`ege, Belgium }

\author{Jens HJORTH}
\affil{Astronomical Observatory, Copenhagen, Denmark  }

\author{Laurence JONES}
\affil{School of Physics and Astronomy, University of Birmingham,
UK}

\author{Olivier LE FEVRE}
\affil{Laboratoire d'Astrophysique, Marseille, France}

\author{Carol LONSDALE }
\affil{Infrared Processing and Analysis Center, Caltech, US }

\author{Dario MACCAGNI}
\affil{IASF, Milano, Italy}

\author{Alain MAZURE}
\affil{Laboratoire d'Astrophysique, Marseille, France }

\author{Brian McBREEN}
\affil{Physics Department, University College, Dublin, Ireland}

\author{Henry McCRACKEN\altaffilmark{5}}
\affil{Laboratoire d'Astrophysique, Marseille, France}
\altaffiltext{5}{currently at  Institut d'Astrophysique, Paris,
France}

\author{Yannick MELLIER}
\affil{Institut d'Astrophysique, Paris, France}

\author{Trevor PONMAN}
\affil{School of Physics and Astronomy, University of Birmingham,
UK}

\author{Hernan QUINTANA}
\affil{Pontificia Universidad Cat\'olica, Santiago, Chile}

\author{Huub ROTTGERING}
\affil{Leiden Observatory, Leiden, The Netherlands}

\author{Alain SMETTE, Jean SURDEJ}
\affil{Institut d'Astrophysique et de G\'eophysique, Universit\'e
de Li\`ege, Belgium}

\author{Jean-Luc STARCK, Laurent VIGROUX }
\affil{CEA/DSM/DAPNIA, Service d'Astrophysique, Saclay, France }

\author{Simon WHITE }
\affil{Max Planck Institut f\"ur Astrophysik, Garching bei
M\"unchen, Germany. }

\newpage

        \begin{abstract}

The XMM Large Scale Structure survey (XMM-LSS) is a medium deep
large area X-ray survey. Its goal is to
   extend  large scale structure investigations attempted using ROSAT
  cluster samples to two redshift bins between $0<z<1$ while
  maintaining the precision of earlier studies. Two main goals have
  constrained the survey design: the evolutionary study of the
  cluster-cluster correlation function and of the cluster number
  density. The adopted   observing configuration consists of an
  equatorial mosaic of 10 ks pointings, separated by $20\arcmin$ and
  covering $8^\circ \times 8^\circ$, giving a point source sensitivity of $\sim
  5~10^{-15}$ erg~cm$^{-2}$~s$^{-1}$ in the [0.5-2] keV band. This will
  yield more than 800 clusters of galaxies and a sample of X-ray AGN
  with a space density of about 300 deg$^{-2}$. We present the
  expected cosmological implications of the survey in the
  context of $\Lambda$CDM models and cluster evolution.
  We give an overview of  the first observational results.

  The XMM-LSS
  survey is associated with several other major surveys, ranging from
  the UV to the radio wavebands which will provide the necessary
  resources for X-ray source identification and further statistical
  studies. In particular, the associated CFHTLS weak lensing and AMiBA
  Sunyaev-Zel'dovich surveys over the entire XMM-LSS area  will
  provide for the first time a comprehensive study of the mass distribution
  and of cluster physics in the universe on scales of a few
  hundred Mpc. We describe the main characteristics of our wavelet-based
  X-ray pipeline and source identification procedures, including the
  classification of the cluster candidates by means of a photometric
  redshift analysis. This permits the selection of suitable targets
  for spectroscopic follow-up. We present  preliminary results
  from the first 25 XMM-LSS pointings : X-ray source
  properties, optical counterparts,highlights from the first Magellan
  and VLT/FORS2 spectroscopic runs as well as preliminary results from
  the NIR search for $z>1$ clusters. The results are promising and, so
  far, in accordance with our predictions. In particular: (1) we
  reproduce the LogN-LogS distribution for point sources obtained from
  deeper surveys at our  sensitivity; (2) we find a cluster number density of 15-20 per
  \dd;
  (3) for the first time, we statistically sample the group mass
  regime at a redshift out to $\sim 0.5 $.

\end{abstract}

\section{Introduction}
\subsection{The role of clusters in cosmology}

We have come a long way since the Virgo cluster of galaxies was
discovered as an accumulation of ``nebulae'' by Messier in 1781,
the first time the term ``cluster'' was used by Shapley \& Ames in
1926, and the first inference of large amounts of ``dark matter''
in the Virgo cluster by Zwicky in 1933, to the point where
clusters of galaxies are now routinely used as cosmological tools.
This progress, is intimately related to the development and
success of the standard cosmological model over the past 50 years,
following ever-deeper insights into the early evolution of the
Universe and the growing power of numerical simulations. From a
simple theoretical point of view, clusters of galaxies - the most
massive bound structures in the Universe - are objects having a
mass of the order of $10^{14-16}~ M_\odot$ growing by accretion at
a rate governed by the initial density fluctuation spectrum, the
cosmological parameters, the nature and amount of dark matter as
well as the nature of the dark energy. Their 3-dimensional space
distribution and number density as functions of cosmic time
constrain cosmological parameters in a unique way.
Non-gravitational effects accompanying cluster formation render
the picture more complicated, but compared to galaxies, clusters
still offer considerable advantages for large scale structure
(LSS) studies: they can provide complete samples of objects over a
very large volume of space, and they are in crucial respects
simpler to understand. The extent (and mass) of clusters can be
traced by their X-ray emission while the theory describing their
formation (biasing) and evolution from the initial fluctuations
can be tested with numerical simulations. Such a level of
understanding does not exist for galaxies - which have reached a
highly non-linear stage -   and even less for QSO and AGN
formation. The resulting cluster LSS counts   can constrain
cosmological parameters, independently of Cosmic Microwave
Background (CMB) and supernova (SN) studies since they do not rely
on the same physical processes. In addition, they can also be used
to test fundamental assumptions of the standard paradigm, such as
the gravitational instability scenario. A quantitative overview of
the cosmological implications of cluster surveys can be found for
instance in \citet{hai}.

Conversely, given a cosmological model, a large and deep
statistical sample of clusters would provide long-awaited
information linking cluster physics, non-linear phenomena involved
in cluster evolution, and scaling relations. Further, cluster
number counts as a function of both the redshift and X-ray
luminosity (or any observable), provide strong consistency tests
on assumptions made in modelling the mass-luminosity (or
mass-observable) relation and cosmological models involving dark
energy \citep{hu}. This is  timely, since following the
WMAP\footnote{http://lambda.gsfc.nasa.gov/product/map/} results,
independent cosmological constraints from both the early and the
local universe, must be integrated into a consistent framework.

\subsection{The quest for clusters}

Initiated by Abell in 1958 over the whole sky in the optical, the
systematic search for clusters underwent a boost of activity in
the X-ray waveband during the EINSTEIN (1978-1981) era. The first
X-ray rockets and satellites had revealed the existence of X-ray
emission associated with clusters. The UHURU (1970) data, in
particular, showed that this emission is thermal and originates in
a hot diffuse gas trapped in the gravitational potential of the
cluster. The Extended EINSTEIN Medium Sensitivity Survey
(EMSS\footnote{http://xml.gsfc.nasa.gov/archive/catalogs/9/9015/})
provided some 730 serendipitous X-ray sources extracted from
pointed observations down to a sensitivity of $1.5~10^{-13}$
erg~cm$^{-2}$~s$^{-1}$ in the [0.3-3.5] keV band. A sub-sample of
67 sources identified as clusters of galaxies in the $0.14<z<0.6$
range, \citep{emss} suggested, for the first time, a mild
evolution in the cluster number density. Next, the ROSAT
All-Sky-Survey\footnote{http://heasarc.gsfc.nasa.gov/docs/rosat/survey/.
In survey mode the PSF HPD is $\sim 130\arcsec $ at 1 keV (from
EXSAS command {\tt calc/psf})} (RASS, 1990-1991) provided the
first sample of X-ray clusters over the entire sky and thus, a
fundamental resource for LSS studies. Cosmological implications
have been investigated in detail from the southern REFLEX ($3
~10^{-12}$ erg~cm$^{-2}$~s$^{-1}$ in the [0.2-2.4] keV band) and
North Ecliptic Pole samples (down to $3~ 10^{-14}$
erg~cm$^{-2}$~s$^{-1}$) by \citet{reflex} and \citet{nep}
respectively.

\begin{figure}[ht]
  \plotone{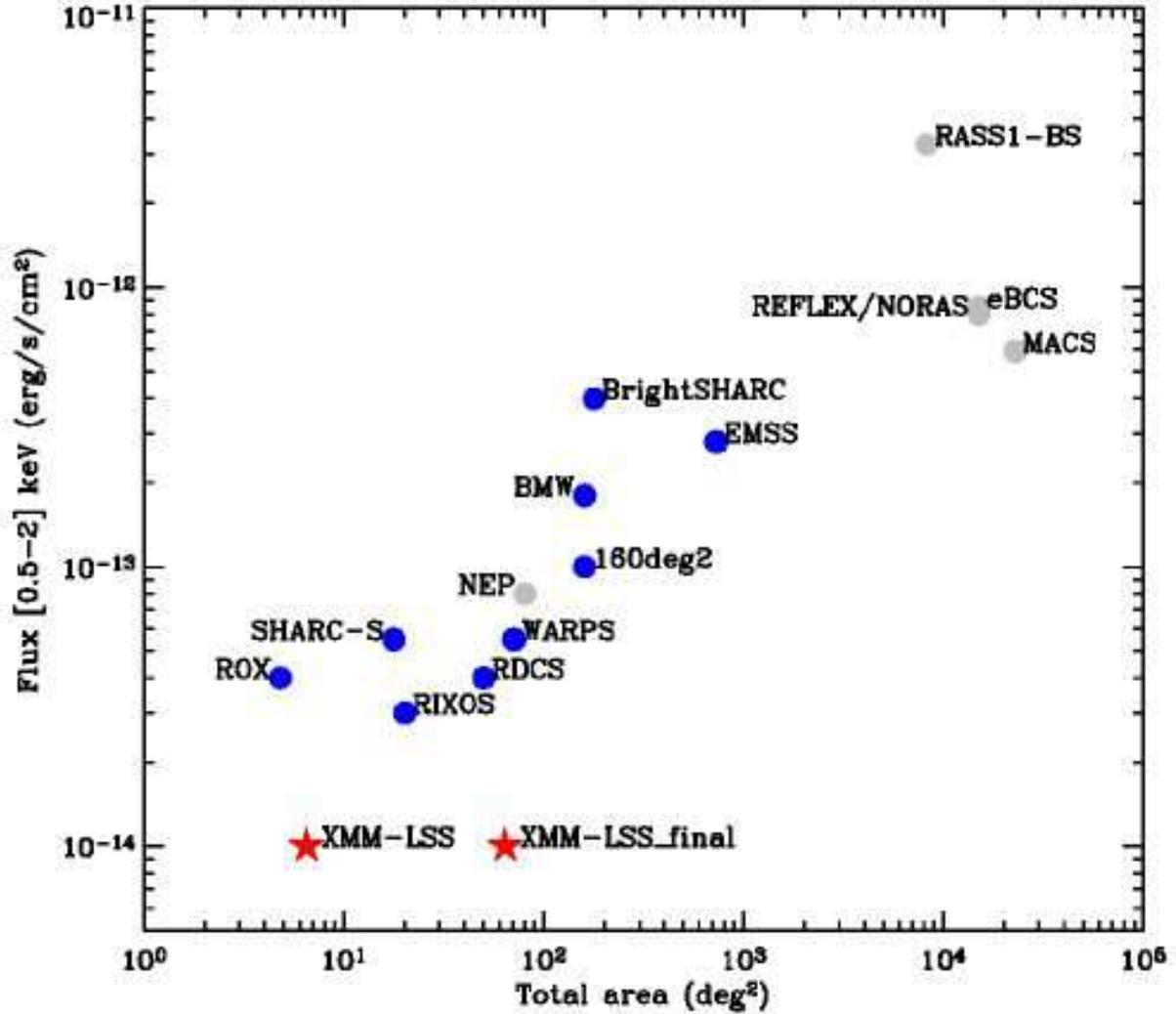}
  \caption[]{Overview of existing X-ray cluster surveys as a function
    of area and  flux. The area plotted is the maximum
area of each survey; the flux plotted is the flux at which the
survey area is half the maximum. The light filled circles indicate
    surveys covering contiguous area, while the blue circles represent
    serendipitous surveys; the stars show the position of the survey
    presented in this paper, the XMM-LSS survey (current stage and the
    foreseen final coverage). References: MACS \citep{macs}, RASS1-BS
    \citep{rass-bs},  REFLEX/NORAS \citep{noras}, \citep{reflex} , EMSS \citep{emss}, , BMW
    \citep{bmw}, Bright SHARC \citep{bsharc}, 160\dd\
    \citep{cfa160}, NEP \citep{nep}, WARPS \citep{warps}, RDCS
    \citep{rdcs}, RIXOS \citep{rixos}, SHARC-S \citep{sharcs}, ROX
    \citep{rox}. Other cluster surveys are CIZA \citep{ciza},  XCS
\citep{xcs} (relevant information not available). For eBCS
\citep{ebcs}, the shape of the flux/area curve was assumed to be
similar to that of REFLEX ). \label{surveys}}
\end{figure}

In parallel, following on the EMSS achievement, serendipitous
searches for distant clusters in deep pointed ROSAT observations
led to the discovery of clusters out to $z\sim 1.2$ with a modest
evolution of the   high luminosity  cluster population; a summary
of the main X-ray cluster surveys is presented on Fig.
\ref{surveys}.

Both, structure and luminosity evolution studies,  are consistent
with hierarchical models of structure formation in a flat low
density universe with matter density $\Omega_{m} \sim 0.3$ and
amplitude of mass fluctuations on 8 h$^{-1}$ Mpc scale,
$\sigma_{8} \sim0.7-0.8$ (see \citet{ros02}, for a review). In
parallel, the quest for clusters in the optical made its own way,
from automated searches on digitised sky survey plates to deep
multi-colour CCD imaging. Indeed, beyond $z\geq 0.8$, detecting
clusters only from galaxy overdensity in a single band is severely
hampered by the faint background population, so that the use of
fine-tuned photometric redshift information becomes mandatory (see
e.g. Sec. 5.1). However, given the limitations on the accuracy of
such methods and various underlying hypotheses about galaxy
evolution, this usually yields large numbers of high-z candidates,
many of them simply being portions of cosmic filaments seen in
projection. One will thus always require in default of extensive
optical spectroscopic campaigns, an ultimate confirmation from the
X-ray band,  to assess the presence of deep potential wells.

\subsection{The power of XMM}

At the time the XMM project was initiated\footnote{first
discussions proposing   27 telescopes in 1982; acceptance in
1988}, the case for surveys was not as compelling as it is today.
Ten years after the completion of the RASS, and following
considerable steps forward in our knowledge of cluster physics
(e.g. \citet{pet}), XMM is in a position to open a new era for
X-ray surveys. Its high sensitivity, considerably better PSF than
the RASS (FWHM $\sim 6\arcsec$ on axis\footnote{the XMM PSF
broadens and becomes irregular  with increasing off-axis angle;
for instance, at 1.5 keV, the radius at which 90\% of the energy
is encircled is $\sim$ 50\arcsec\ and 65\arcsec\ for off-axis
angles of 0 and 12\arcmin\ respectively for the pn detector; HPD
is 15\arcsec\ on-axis at 1.5 keV (cf XMM Users Handbook)}) and
large field of view ($30\arcmin$), make it a powerful tool for the
study of extragalactic LSS. In this respect, two key points may be
emphasised. Firstly, a high galactic latitude field observed with
XMM at medium sensitivity ($\sim 0.5-1~ 10^{-14}$
erg~cm$^{-2}$~s$^{-1}$) is ``clean'' as it contains only two types
of objects, namely QSOs (pointlike sources) and clusters (extended
sources) well above the confusion limit. Secondly, if clusters
more luminous than $L_{[2-10]} \sim 3 ~10^{44}$ h$_{70}^{-2}$erg/s
are present at high redshift, they can be detected as extended
sources out to $z = 2$, in XMM exposures of only 10 ks. XMM is a
powerful wide angle X-ray imager, with a sensitivity to extended
sources which will remain unrivalled in the coming decade. In
parallel, progress in optical wide field imaging makes it possible
to undertake optical identification of faint X-ray sources over
tens of square degrees. Consequently, we are now in a position to
probe the evolution of the cosmic network traced by clusters and
QSOs over large volumes of the Universe to high redshift.

\subsection{The goals of the XMM-LSS}

Each new generation of instruments brings  major improvements over
its predecessors. While the BCS sample presented hints of
structure in the sky distribution of X-ray bright clusters, REFLEX
was the first X-ray survey to systematically address LSS with
clusters in the nearby universe (450 clusters with $z\leq 0.2$).
Ten years later, there is  a clear need to investigate the
evolution of the cosmic network out to $z \sim 1$. In this
context, we have designed a survey to yield some 800 clusters in
two redshift bins with $0<z<1$: the XMM Large Scale Structure
Survey (XMM-LSS). This simple goal has set the sensitivity and the
coverage of the XMM-LSS and, as shown in Sec.2.1, implies an X-ray
sensitivity about two orders of magnitude deeper than REFLEX.
Consequently, this survey will determine how the cluster number
density evolves - a hotly debated topic. It will also trace the
LSS as defined by X-ray QSOs out to redshifts of $\sim 4$. The
cosmological implications of the XMM-LSS are summarised in Sec. 2.
In addition, the proposed X-ray survey is associated with several
other major new   surveys (optical, IR, Radio, UV). This will
provide a new data set that can be used to study the evolution of
clusters, cluster galaxies, and   star and AGN formation as a
function of environment, in the context of structure formation. An
overview of these capabilities is presented in Sec. 3. In the
following sections we present a summary of results derived from
the XMM AO-1 data set, demonstrating the feasibility of the
project. Specifically, in Sec. 4, we describe our current X-ray
source identification procedure, which led to the first
spectroscopic and NIR follow-up campaigns in Autumn 2002 (Sec. 5
\& 6). Future activities and immediate improvements are presented
in Sec. 7.

\subsection{The XMM-LSS consortium}

The wide scope of the project has motivated the assembly of a
large consortium to facilitate both the data reduction/management
and the scientific analysis of the survey. The XMM-LSS Consortium
comprises some fifteen European and Chilean institutes.  The
project is presented in detail on the following website: {\tt
http://vela.astro.ulg.ac.be/themes/spatial/xmm/\\LSS/index\_e.html}

\begin{figure}[ht]
  \centerline{
    \includegraphics[height=8cm,angle=-90]{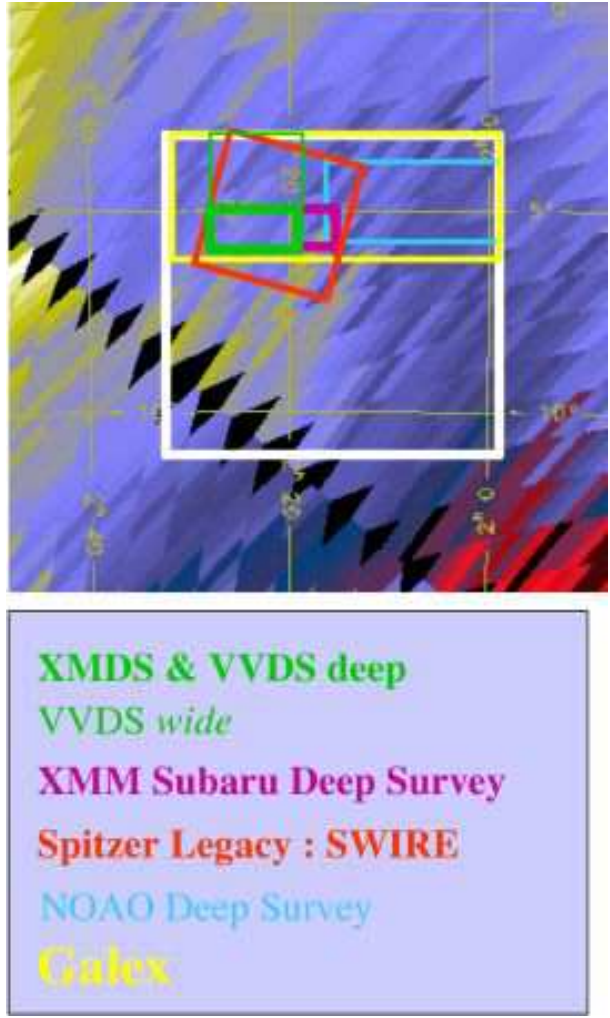}
  }
  \caption[]{ Large white square indicating the location of the
    XMM-LSS survey is overlaid on a map of
   $N_H$ ($1.4~10^{20}<N_H (cm^{-2})<3.5~10^{20}$ for the XMM-LSS field). The
    survey area surrounds two deep XMM surveys based on guaranteed
    time: the XMM\_SSC/Subaru Deep Survey (80~ks exposures in $1 \ \rm
    $\dd) and the XMM Medium Deep Survey (XMDS; 20~ks exposures in $2
    \ \rm $\dd) also corresponding to the VIRMOS-DESCART Deep
    Survey[$deep$], the latter being a collaboration between several
    instrumental teams: XMM-OM (Li\`ege), XMM-EPIC
    (IASF-MILANO),XMM-SSC (Saclay); CFHTLS (Saclay, IAP); VIRMOS (LAM,
    IASF-MILANO,OAB).  Also indicated  are
    the positions of the associated DESCART-VIRMOS Deep Survey
    [$wide$], the SWIRE SPITZER Legacy Survey, the Galex survey and the NOAO deep survey
    (Sec. 3); the 8.75 \dd\ UKIDSS survey is centered on the Subaru Deep
    Survey and the CTIO R-z' imaging covers a region corresponding to
    the VVDS [$wide$] and [$deep$] surveys.(Note that the center of the whole XMM-LSS survey has
    been shifted by 2 degrees southward from its initial position
    because of the presence of the variable type M7 star Mira Ceti (02
    19 20.8-02 58 39) which can reach I=1,  and cause difficulties
    with optical mapping.) \label{map}}
\end{figure}

\section{Survey design and cosmological implications}
\subsection{Defining the survey}

Given our driving goal of extending the REFLEX achievements to
high redshift (Sec. 1.4), the following objectives drove the
design of the survey.

\noindent$\bullet$ {\bf Measure the cluster correlation function
in two redshift bins spanning $ 0 < z < 1$, with a good level of
accuracy.} A first estimate using the Hubble Volume
Simulations\footnote{http://www.mpa-garching.mpg.de/Virgo/hubble.html}
of a $\Lambda$CDM universe, showed that some 400 clusters per
redshift bin are necessary in order to obtain  an accuracy of
15-20\% on the correlation length; this is comparable to the size
of the low-z REFLEX sample. Subsequent implications for the
accuracy of the determination of the cosmological parameters are
summarized in Sec. 2.2.

\noindent$\bullet$ {\bf Probe a comoving length which is
significantly larger than 100 h$^{-1}$Mpc at ~$ z \sim 1$, the
characteristic scale in the galaxy power spectrum of the local
universe (e.g. \citet{lan}).} This constraint corresponds to an
opening survey angle of $\sim10^\circ$ at $z = 1$ (i.e. 400
h$^{-1}$Mpc)

\noindent$\bullet$ { \bf Find the best compromise between the two
above constraints in order to minimise the necessary XMM observing
time.} Quantitatively, we used the Press-Schechter formalism
\citep{ps74} and the mass-temperature relation from simulations to
predict the counts of clusters and their X-ray properties in
several CDM models \citep{ref}. We computed the detection
efficiency of clusters, using realistic simulations of XMM images,
and studied how this differs from a conventional flux limit.

In order to fulfil the first 3 conditions above, and assuming the
current favoured $\Lambda$CDM cosmological model, the optimal
survey design was found to be an $8^\circ\times 8^\circ$ area,
paved with 10 ks XMM pointings separated by 20 arcmin (i.e. 9
pointings per \dd). The expected ultimate sensitivity is $\sim
3~10^{-15}$~erg~cm$^{-2}$~s$^{-1}$ for pointlike sources in the
[0.5-2] keV band. This is about 1000 and 10 times deeper than the
REFLEX \citep{reflex} and NEP \citep{nep} single area surveys
respectively and provides much better angular resolution
($6\arcsec$ vs $2\arcmin$). So far, some 6 \dd\ have been covered
by XMM in the region.

\noindent$\bullet$ {\bf Find an adequate survey location.}  An
equatorial field is optimal, as ground-based follow-up resources
from both hemispheres may be used. High galactic latitude and the
absence of bright X-ray sources (e.g. nearby clusters) are also
required. Moreover, the visibility of the field by XMM must be
$\geq 15\%$. Given this, only one area in the sky turned out to be
favorable: a field centred around $\rm \alpha =
2^h18^m00^s$,$\delta=-7^\circ 00'00''$ (at $b=-58^\circ$) with
neutral hydrogen column $1.4 \times 10^{20} < N_H ({\rm cm^{-2}})
< 3.5 \times 10^{20}$. Whereas this region appears to be the best
compromise between many astronomical and instrumental constraints,
it should be noted that it is not optimal for far infra-red
observations (above 25 $\mu$m) because of cirrus contamination.
The X-ray survey is centred on two deeper areas (one and two
square degrees) deriving from XMM guaranteed time programmes (GT).
Several surveys are located in the same region (Sec.3) of which an
overview map is presented on Fig. \ref{map}.

\medskip

\begin{figure*}[ht]
  \centerline{\includegraphics[height=20cm,angle=-90]{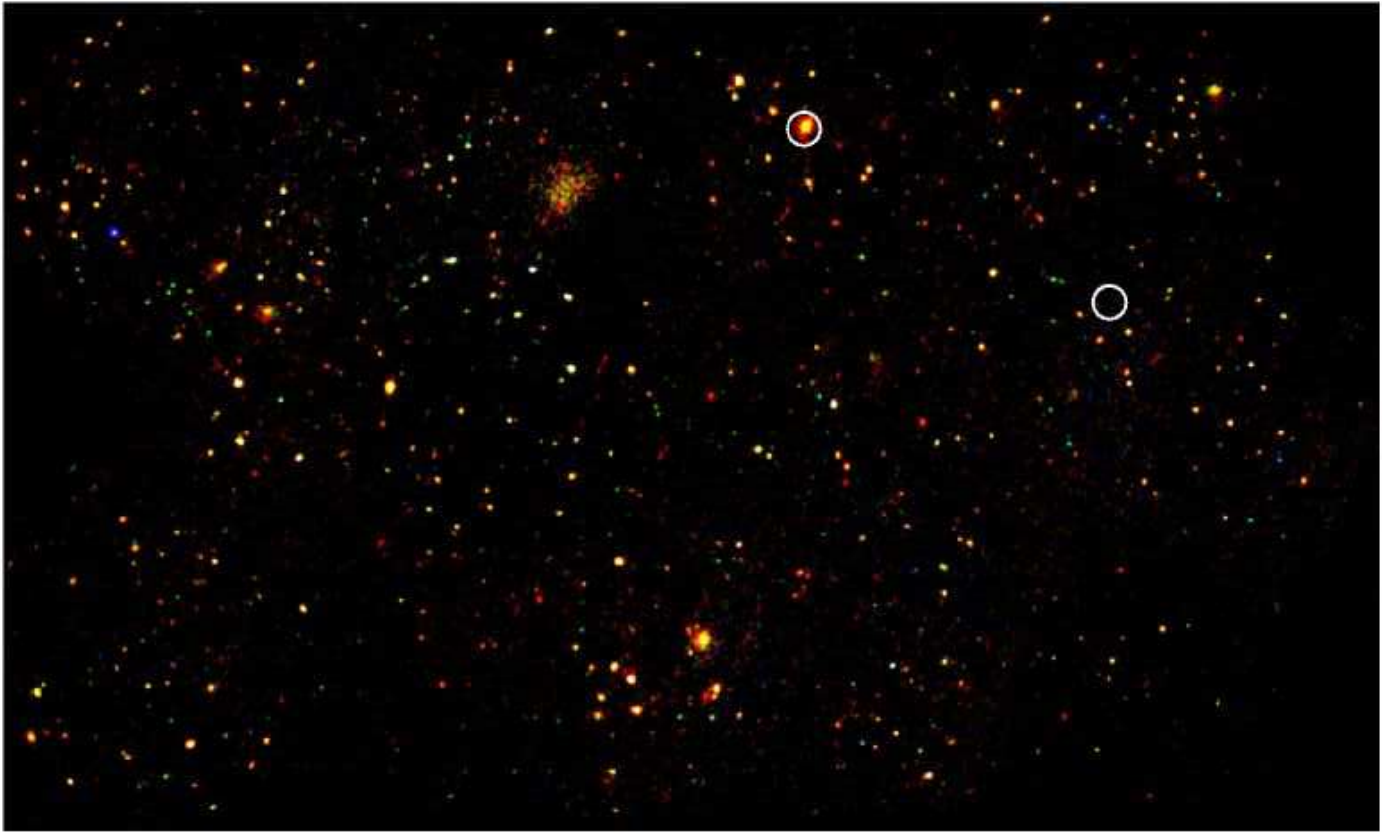}}
  \caption[]{{\bf First view of the deep X-ray sky on large
      scales}. Image obtained combining the first 15 XMM-LSS fields
      mosaiced in true X-ray colours: red [0.3-1.0] keV, green
      [1.0-2.5] keV, blue [2.5-10.0] keV. The circles indicate the
      sources found in the RASS; the brightest one being a star,
      HD14938. The other source, apparently not present in our XMM data,
      is from the RASS Faint Source Catalogue with only 7 counts, and thus
      probably a spurious detection. The displayed region covers 1.6 \dd.
      This is the first
      time that such an X-ray depth has been achieved over such an
      area. The improvement with respect to the RASS is striking, with
      a source density of the order of $\sim$ 300~deg$^{-2}$ in the
      [0.5-2] keV band. The wealth of sources includes supersoft and
      very hard sources, as well as sources with a wide range of
      intrinsic extents, giving an indication of the scientific
      potential of the XMM-LSS survey.}\label{mosaic}
\end{figure*}

\subsection{Constraining cosmology - Quantitative assessments}
The prospects that the XMM-LSS cluster catalogue offers for
determining cosmological parameters have been studied in detail by
\citet{ref}. We recall here the main issues.

$\bullet$ The cluster counts set strong constraints on the value
of the ($\Omega_{m}-\sigma_{8}$) combination. This combination
will also provide a consistency check for the $\Lambda$CDM model,
and a discrimination between this model and the OCDM model.

$\bullet$ The addition of the cluster 2-point correlation function
provides a constraint on $\Gamma$, the shape of the initial
density fluctuation power spectrum.

$\bullet$ With the current survey design, the $simultaneous$
expected precision on $\Omega_{m}, \sigma_{8}$ and $\Gamma$ is
about 15\%, 10\%, 35\% respectively. It is important to note that
the present uncertainties on $\sigma_{8}$ globally result in a
factor of 2 uncertainty in our predicted cluster numbers of
$\sim$15~deg$^{-2}$ (i.e. from 600 to 1200 cluster detections
expected within $0 < z < 1$). The high sensitivity to $\sigma_{8}$
is not surprising, as it is precisely this which makes cluster
counts a good measure of this parameter. This uncertainty can be
reduced by analysing about 15 \dd\ of the XMM-LSS, the minimum
area required to improve upon the current measurements of
$\sigma_{8}$ in the presence of shot noise and cosmic variance.

$\bullet$ The sensitivity of the XMM-LSS survey allows the entire
cluster population ($\geq 2$ keV) to be detected out to a redshift
of 0.6, and will unveil the nearby group population (the cluster
selection function can be found in \citet{ref}). Rather few galaxy
groups have been well-studied in the nearby universe, whereas they
are believed to constitute the majority of the mass, and of the
baryon reservoir. They are also the building block of richer
clusters. With increasing redshift, the XMM-LSS is less sensitive
to low mass systems. Consequently, the low-z and high-z samples to
be used for the study of the LSS, will pertain to different
cluster mass ranges. But as outlined above, this does not prevent
the derivation of strong cosmological constraints. The XMM-LSS
high-z clustering properties will be compared to the REFLEX ones,
which accurately sample the most massive nearby clusters.
Moreover, given the large volume sampled at high redshift, the
XMM-LSS is well suited to constraining the abundance of distant
massive clusters. For instance, the probability of finding a
Coma-like cluster in the $1.5<z<2$ range over 64 \dd\ is the order
of $ 6~ 10^{-7}$ for a $\Lambda$CDM universe. Finding a few such
objects would thus put the current favored cosmological model into
severe troubles, or strongly question our understanding of cluster
physics at high redshift.

\section{Associated multi-wavelength surveys : an overview}
While optical information remains the primary data base for X-ray
source identification, contribution from other wave bands may be
critical, for example in the far infra-red domain, where many
heavily absorbed X-ray QSO, not visible in the optical, are
expected to show up. Beyond the necessary identification step,
multi-wavelength information provides an overview of the energy
emission and absorption processes in astronomical objects, which
is vital to our understanding of their physics, formation and
evolution. The contiguous design of the XMM-LSS, optimised for
large scale studies, provides considerable advantages for
complementary observations, compared to serendipitous fields, and
the project has developed numerous collaborations at other
wavelengths. The main characteristics of associated surveys are
summarised in Table \ref{flwp} and their main science applications
are outlined below.

\begin{table*}
\caption[]{\bf XMM-LSS X-ray and associated surveys. \label{flwp}}
\small
\begin{tabular}{ l l l l }
  \tableline\tableline
  Observatory/Instrument & (Planned) Coverage & Band & Final
  Sensitivity \\
  \tableline
  XMM/EPIC & 64 \dd & [0.2-10] keV & $\sim 5-3~10^{-15}$erg~cm$^{-2}$~s$^{-1}$ [1] \\
  CFHT/CFH12K (VVDS Deep) *& 2 \dd\ GT & B, V, R, I & 26.5, 26.0, 26.0, 25.4 [2] \\
  CFHT/CFH12K (VVDS Wide) *& 3 \dd\ GO & V, R, I & 25.4, 25.4, 24.8 [2] \\
  CFHT/MegaCam & 72 \dd & u'', g', r', i', z' & 25.5, 26.8,26.0, 25.3, 24.3 [3] \\
  CTIO 4m/Mosaic & $\sim$ 16 \dd & R, z' & 25, 23.5 [4] \\
  UKIRT/WFCAM & 8.75 \dd & J, H, K& 22.5, 22.0, 21.0 [5] \\
  VLA/A-array *& 110 \dd & 74 MHz & 275 mJy/beam [6a]\\
  VLA/A-array & 5.6 \dd & 325 MHz & 4 mJy/beam [6b]\\
  OCRA & all XMM-LSS clusters & 30 GHz & 100 $\mu$Jy [7]\\
  AMiBA & 70 \dd & 95 GHz & 3.0 mJy [8] \\
  SPITZER/IRAC (SWIRE Legacy)* & 8.7 \dd & 3.6, 4.5, 5.8, 8.0 $\mu$m &
  7.3, 9.7,27.5, 32.5 $\mu$Jy [9a]\\
  SPITZER/MIPS (SWIRE Legacy)* & 8.9 \dd & 24, 70, 160 $\mu$m & 0.45,
  6.3, 60 mJy [9b]\\
  Galex & $\sim 20$ \dd & 1305-3000 \AA & $\sim 25.5$ [10] \\
\hline
\end{tabular}
 Notes:

    * :complete

    [1]: for pointlike sources in [0.5-2] keV

    [2]: AB$_{Mag}$, $5\arcsec$ aperture

    [3]: S/N = 5 in $1.15\arcsec$ aperture

    [4]: 4 $\sigma$ in $3\arcsec$ aperture

    [5]: Vega$_{Mag}$

    [6a]: $30\arcsec$ resolution; deeper observations
    planned

    [6b]: $6.3\arcsec$ resolution

    [7]: $5 \sigma$, detection limit

    [8]: $6 \sigma$, detection limit

    [9a]: $5 \sigma$

    [9b] $5 \sigma$

    [10]: AB$_{Mag}$

\end{table*}

\underline{Optical, NIR and UV imaging:} The imaging of the
$8\times 8 $ \dd\ XMM-LSS area is one of the priorities of the
Canada-France-Hawaii Legacy
Survey\footnote{http://cdsweb.u-strasbg.fr:2001/Science/CFHLS/cfhtlscfhtoverview.html}
(CFHTLS). It is being performed by MegaCam, the one degree field
imager built by CEA and installed at the new CFHT prime focus.
Imaging of the XMM-LSS region   started by mid 2003. The CFHTLS
will provide deep high quality optical multi-colour imaging
counterpart of the X-ray sources at a rate of 25 \dd/yr in at
least three colours. Data pipelines and processing have been
developed by the TERAPIX\footnote{http://terapix.iap.fr}
consortium which provides object catalogues and astrometric
positions for the entire surveyed region. Currently, the optical
data used for the identification work, and presented in this
paper, mostly pertain to the CFH12K VIRMOS-DESCART VLT Deep
Surveys (VVDS [$deep$] and [$wide$], Fig. \ref{map} and
Tab.\ref{flwp}, \citep{lef}). An optical cluster catalogue is
under construction  using both spatial clustering analysis and
multi-colour matched filter techniques, in addition to photometric
redshift estimates. Moreover, the MegaCam data will form the basis
of a weak lensing
analysis\footnote{http://www2.iap.fr/LaboEtActivites/ThemesRecherche/\\Lentilles/LentillesTop.html},
whose cosmological constraints will be compared to those provided
by the X-ray data on the same region. This will be the first,
coherent study of LSS on such scales. R and z' imaging taken by us
at CTIO are also being analysed, forming the basis of an
independent cluster catalogue (Sec. 5.1, \citet{and}). In
addition, deep NIR VLT imaging (J \& K ) of $z>1$ cluster
candidates found in the XMM-LSS is performed as a confirmation
prior to spectroscopy. Finally, a sub area of 8.75 \dd\ of the
XMM-LSS field is a high priority target of the Deep Extragalactic
Survey part of the UKIRT Deep Sky Survey
(UKIDSS\footnote{http://www.ukidss.org/}). In the UV domain, the
XMM-LSS field is one of the targets of the
Galex\footnote{http://www.srl.caltech.edu/galextech/galex.htm}
Deep Survey, whose main goal is to map the global history of star
formation out to $z\sim 2$.

\underline{Spectroscopy:} The standard spectroscopic follow-up is
designed to perform redshift measurements for all identified
$0<z<1$ X-ray clusters in Multi-Object-Spectroscopy mode, using 4m
and 8m class telescopes. Current identification procedures and
first results are described in later Sections. We shall
subsequently undertake programmes of advanced spectroscopy that
will focus on individual objects, and include high resolution
spectroscopy, the measurement of cluster velocity dispersions and
QSO absorption line surveys, as well as NIR spectroscopy of our
$z>1$ cluster candidates.

\underline{Radio:} In the radio waveband, the complete survey
region is being mapped using the VLA at 74MHz and 325MHz. First
results of this low frequency coverage are described by
\citet{coh}. Radio observations are not only particularly relevant
for tracing merger events triggered by structure formation, but
also as a useful indicator of galactic nuclear or star-formation
activity.

\underline{Sunyaev-Zel'dovich:} Observations (S-Z) are also
planned. Clusters in the XMM-LSS field are the targets of the
prototype OCRA (One-Centimeter Radiometer Array, \citet{bro})
instrument. The full XMM-LSS field will be mapped by the OCRA
(start  in 2005), and will be an early target of the Array for
Microwave Background Anisotropy (AMiBA) \citep{lo}. This will
enable not only the measurement of the Hubble constant, but also a
statistical analysis of the physics of the ICM as a function of
redshift. For instance, the AMiBA deep survey will be sensitive to
clusters with M $\geq 1.5~10^{14}$ out to $z<2$ \citep{lia}. A
cross-correlation between the AMiBA and XMM maps in the survey
region, will thus allow us to study the gas properties of X-ray
selected clusters and to detect clusters beyond the X-ray
detection threshold. Such measurements are complementary to the
X-ray and weak lensing surveys, connecting the mass distribution
of clusters to the structure of the hot gas they contain. The
three data sets together will also provide a direct and
independent check of the extragalactic distance scale.

\underline{Infrared}: In the infrared, the
SWIRE\footnote{http://www.ipac.caltech.edu/SWIRE} SPITZER Legacy
Programme   covered $\sim 9$ \dd\ of the XMM-LSS in 7 MIR and FIR
wavebands from 4 to 160 $\mu$m \citep{lon}. The estimated IR
source densities per square degree in this area are around
1100/400/130 and 670/150/130 for starbursts/spiral-irregular/AGN
in the $0<z<1$ and $1<z<2$ redshift intervals, respectively
\citep{xu}. This represents a unique X-ray/IR combination in depth
and scales to be probed. The coordinated SWIRE/XMM-LSS
observations will clarify an important aspect of environmental
studies, namely how star formation in cluster galaxies depends on
distance from the cluster centre, on the strength of the
gravitational potential, and on the density of the ICM. In this
respect the XMM-LSS represents the optimum SWIRE field, where
galaxy environment, deep NIR imaging and optical spectroscopic
properties will be the main parameters in modelling the MIR/FIR
activity.  Here also, the location of IR AGNs within the cosmic
web will help establish their nature. The FIR/X/optical/radio
association will also provide valuable insights into the physics
of heavily obscured objects, as well as the first coherent study
of biasing mechanisms as a function of scale and cosmic time, for
hot (XMM), dark (weak lensing), luminous galactic (optical/NIR)
and obscured (SWIRE) material.

In summary, we may underline the complementarity of the two major
categories of associated surveys. Those covering an area of the
order of 10-20 \dd, focussing on source environment, will provide
new insights on the physics of AGN and clusters. Surveys planned
over a $\sim$ 70 \dd\ area  (MegaCam and AMiBA), in addition to
proper LSS studies, are intended strong consistency tests on
assumptions made in modelling the cluster mass/luminosity and
mass/temperature relations and cosmological models involving dark
energy.

\begin{figure*}[ht]
  \centerline{\includegraphics[width=17cm,angle=0]{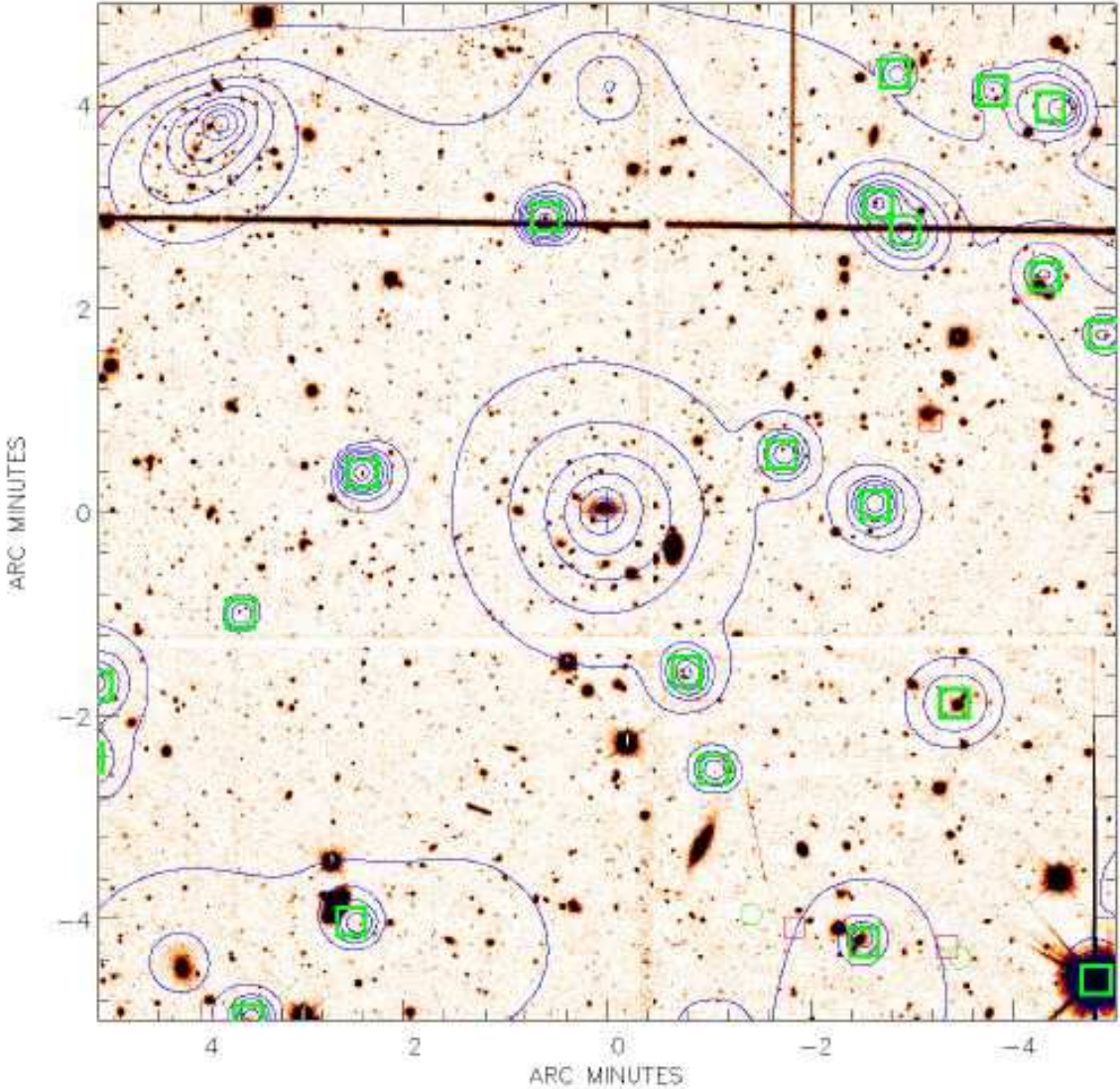}}
  \caption[]{This image ($10 \arcmin \times 10 \arcmin$) provides an
    overview of the source types encountered in the survey and of the
    identification procedure. Overlaid on a 1h exposure CFH12KI-band
    image (processed by Terapix) are blue XMM flux contours in [0.5-2]
    keV, obtained by the MR1 multi-scale wavelet filtering algorithm;
    the procedure allows the automatic recognition of pointlike
    sources (green squares); the significance of the lowest wavelet
    contour is $10^{-4}$ (equivalent to $3.7 \sigma$ for Gaussian
    noise, cf \citet{sta}).  The image is centred on a bright extended
    X-ray source, corresponding to an obvious nearby cluster (a {\em
    NEAR} candidate) for which we measured a redshift of 0.33. Another
    - much more distant - cluster is apparent in the upper left corner
    (a {\em MID} candidate); this object was found to have a redshift
    of 0.84. Top middle, another extended source without a clear
    optical counterpart, typically a {\em DISTANT} candidate. The
    density of pointlike sources is high, some of them having an
    obvious optical counterpart, others, none. The red squares and
    green circles indicate VLA NVSS and 325 MHz radio sources
    respectively. In the lower right corner, there is a conspicuous
    double lobe radio galaxy, which is also an X-ray source.}
    \label{cluster1}
\end{figure*}

\section{X-ray source lists}
 In addition to the  XMDS Guaranteed Time survey consisting of 18
pointings (Fig. \ref{map}), 33 Guest Observer XMM-LSS pointings
have been allocated in the first two XMM AOs.  The results
presented in this section pertain to the first 27 GO XMM-LSS
pointings allocated in AO1. Most of the X-ray observations were
performed in  good background conditions. However, two pointed GO
observations (namely B13 and B17) are heavily contaminated by
solar flares, having a mean total count-rate approximately 100
times higher than the quiescent background. Effective exposure
times range from 5 to 16 ks (after removal of bad time intervals,
mainly due to flares). The mean exposure time weighted over the
area is about 12 ks and 9 ks for the MOS and pn detectors
respectively, which is close to the nominal survey exposure time
of 10 ks. Statistics  presented in this section are obtained using
the 25 good pointed observations that were used out to an off-axis
angle of 10\arcmin ~and cover a total area of 2.18 \dd.

\subsection{ Source detection and statistics}
The XMM-LSS pipeline is based on a 3 stage
filtering/detection/measurement process  initiated by
\citet{val01}. After applying the standard XMM pre-reduction
procedure\footnote{
http://xmm.vilspa.esa.es/external/xmm\_sw\_cal/sas\_frame.shtml},
photon images are generated in several energy bands, and  the 3
detector images are co-added  in  each band.

\begin{figure}[ht]
\centerline{\includegraphics[width=10cm,angle=90]{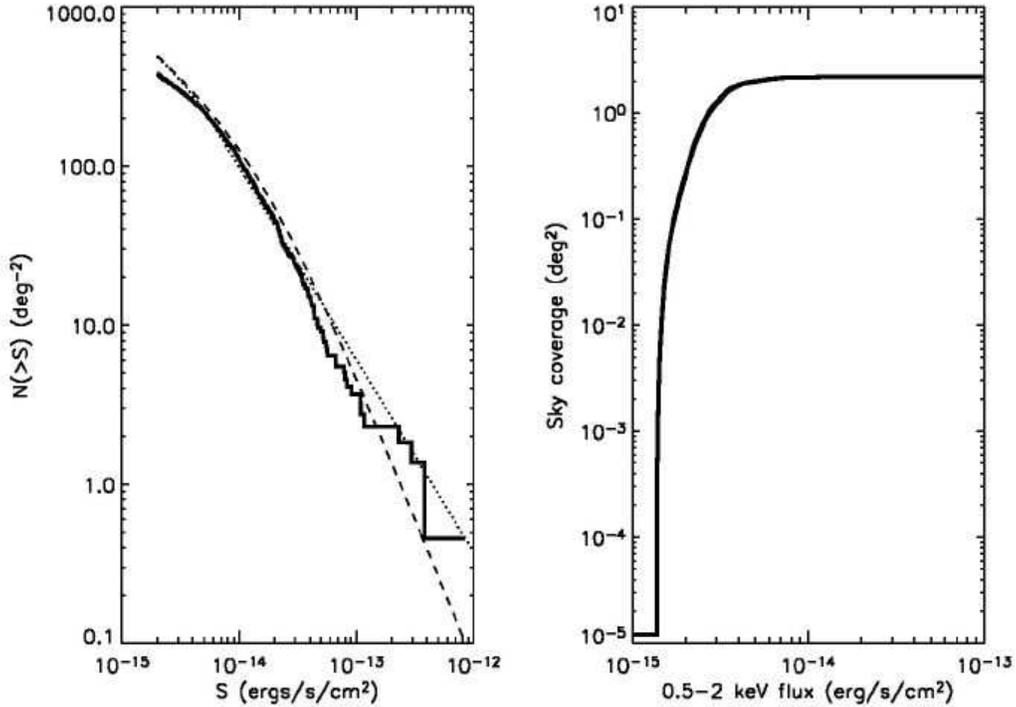}} \caption[]{ Left: Cumulative log\,N-log\,S distribution, in the
[0.5-2] keV band, for the sources pertaining to the first 25
XMM-LSS pointings (off-axis $<$ 10 \arcmin, $N_{counts} > 20$ and
detection likelihood $>20$). The dashed line is the analytical fit
of the deep log\,N-log\,S distribution, by \citet{mor}, the dotted
line the determination by \citet{bal} for the HELLAS2 survey.
Right: Corresponding sky coverage for point sources shows that
90\% of the area is covered with a sensitivity better than
$5\times 10^{-15}$~erg~cm$^{-2}$~s$^{-1}$. On-going pipeline
developments, coupled with extensive image simulations, aim to a
deeper completeness limit. \label{lognlogs}}
\end{figure}

This mosaic is subsequently   filtered by a multi-resolution
algorithm (in counts, to preserve Poisson statistics): a
scale-dependent wavelet coefficient threshold, (corresponding to a
probability for each event of being a random background
fluctuation lower than $10^{-3}$) is computed in wavelet space by
histogram auto-convolution, and followed by an iterative
reconstruction of the image (MR1, see \citet{sta}). The source
extraction procedure SExtractor \citep{ber} is then applied on the
resulting image  to obtain a preliminary list of sources. At this
stage, extraction parameters  are set in order to avoid missing
faint sources. Co-adding images allows us to increase the S/N
before filtering and, consequently, to  lower   the detection
thresholds in the first step. Third step consists in examining the
likelihood of the detections and characterizing their extent by
means of a procedure we have developed ({\tt Xamin}, \citet{pac}).
The principle is the following: For each Sextractor source, it
performs a maximum likelihood profile fit on two raw photon images
(PN image, and co-added  MOS images) independently, thus yielding
a quantitative assessment of the source, taking into account the
PSF variation with energy and off-axis radius, as well as other
detector characteristics (vignetting, bad pixels, CCD gaps). Two
source models are tested: a PSF model provided by the XMM
calibration files, and a spherically symmetric $\beta$-model (with
fixed $\beta=2/3$) convolved with the PSF. Models are weighted by
the exposure maps at the location of the fit. The parameters that
are varied for each source are: position, local background, count
rates and core radius of the $\beta$-model. Position and extension
are forced to be the same in each detector, while  count rates are
adjusted independently  for each EPIC detector type.  At the end
of the analysis   {\tt Xamin} yields a catalogue of  some 50
parameters for each source including: source count rates (in each
detector type), likelihood of detection, likelihood of source
extent, and extension in arcsec. Finally, discrimination  between
pointlike and extended sources, is performed in the [Detection
Likelihood / Extent] parameter space, calibrated by means of
extensive image simulations spanning a wide range in source counts
and extent. Performing this analysis on the 25 first  10 ks
pointings, we detect some 200 sources per square degree down to a
flux limit of $5~10^{-15}$~erg~cm$^{-2}$~s$^{-1}$ in the [0.5-2]
keV band for a detection likelihood threshold of 20 (assuming a
power-law spectrum with a photon index of $\Gamma = 1.7$).
Corresponding log\,N-log\,S curve is shown on Fig. \ref{lognlogs}.

\subsection{Extended sources and visual check}
\begin{figure}[ht]
\centerline{\includegraphics[width=8cm,angle=0]{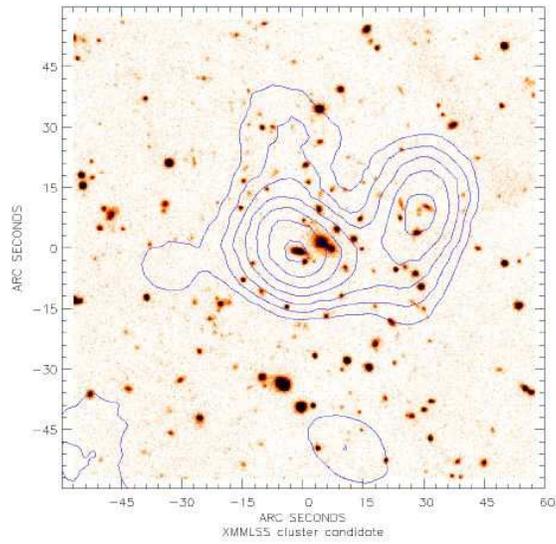}}
\caption[]{The figure shows an XMM  source detected as extended
with only $\sim 50$ net counts in the [0.5-2] keV combined image
of the 3 detectors. The overlay on the optical image confirms the
presence of a distant cluster ($z\sim 0.8-0.9$), and even reveals
X-ray structure associated with 2 distinct groups of galaxies. The
sources actually lies at an off-axis angle of 11.5\arcmin. The
collected photons correspond to an emitted flux of  $\sim
1.2\times 10^{-14}$  or $\sim 4\times 10^{-15}$
~erg~cm$^{-2}$~s$^{-1}$, for a source falling at an off-axis
distance of 11\arcmin\ or on-axis  respectively, assuming a
typical thermal cluster spectrum. \label{limita}}
\end{figure}

As clusters of galaxies constitute the core of the project,
special care is devoted to the detection and assessment of faint
extended sources. The MR1/Sextractor/{\tt Xamin} hybrid method
described above provides a series of parameters used to establish
a preliminary list of extended sources. Currently, the procedure
is performed on the [0.5-2] keV catalogue which offers optimal
sensitivity considering the observed cluster emission spectra over
the $0<z<1$ range, the various components of the background
(galactic, particle, solar flares), and galactic absorption. We
have developed an automated interface to produce X-ray/optical
overlays for every extended source candidate (Fig.
\ref{cluster1}). At the end of the pipeline procedure, each
overlay is inspected by eye in order to catch possible
instrumental artefacts which could have escaped the pipeline
rejection algorithm. Finally, a list of plausible extended sources
is issued with an indication of whether they correspond to an
obvious overdensity of galaxies. We currently find an extended
source density of about 15-20~deg$^{-2}$, with conspicuous optical
cluster (or group) counterparts on the CFH12K I images. At this
stage, given the small number of spectroscopically confirmed
clusters, it is not possible to make firm statements about these
results. The fact that we tend to find somewhat more extended
objects than formally predicted, would suggest that either we
sample the cluster population below the 2 keV limit that was set
in our theoretical predictions, using a given set of M-T-L(z)
relations or that we are more sensitive than assumed. An example
of the sensitivity performances for extended sources is displayed
in Fig. \ref{limita}. This suggests that we reach a sensitivity of
at least $ 10^{-14}$~erg~cm$^{-2}$~s$^{-1}$ in the [0.5-2] keV
band to extended sources of  over a large fraction of the
detector. Which is in agreement with the predictions of
\citet{ref}, showing that the cluster selection function is close
to a flux limit of $ 10^{-14}$~erg~cm$^{-2}$~s$^{-1}$ for $z<1$
clusters. The detailed study of the log\,N-log\,S distribution of
extended sources and of their selection function will be the
subject of a forthcoming paper.

\section{Source classification}
\subsection{Clusters}

\begin{figure}[ht]
  \centerline{\includegraphics[width=6cm,angle=-90]{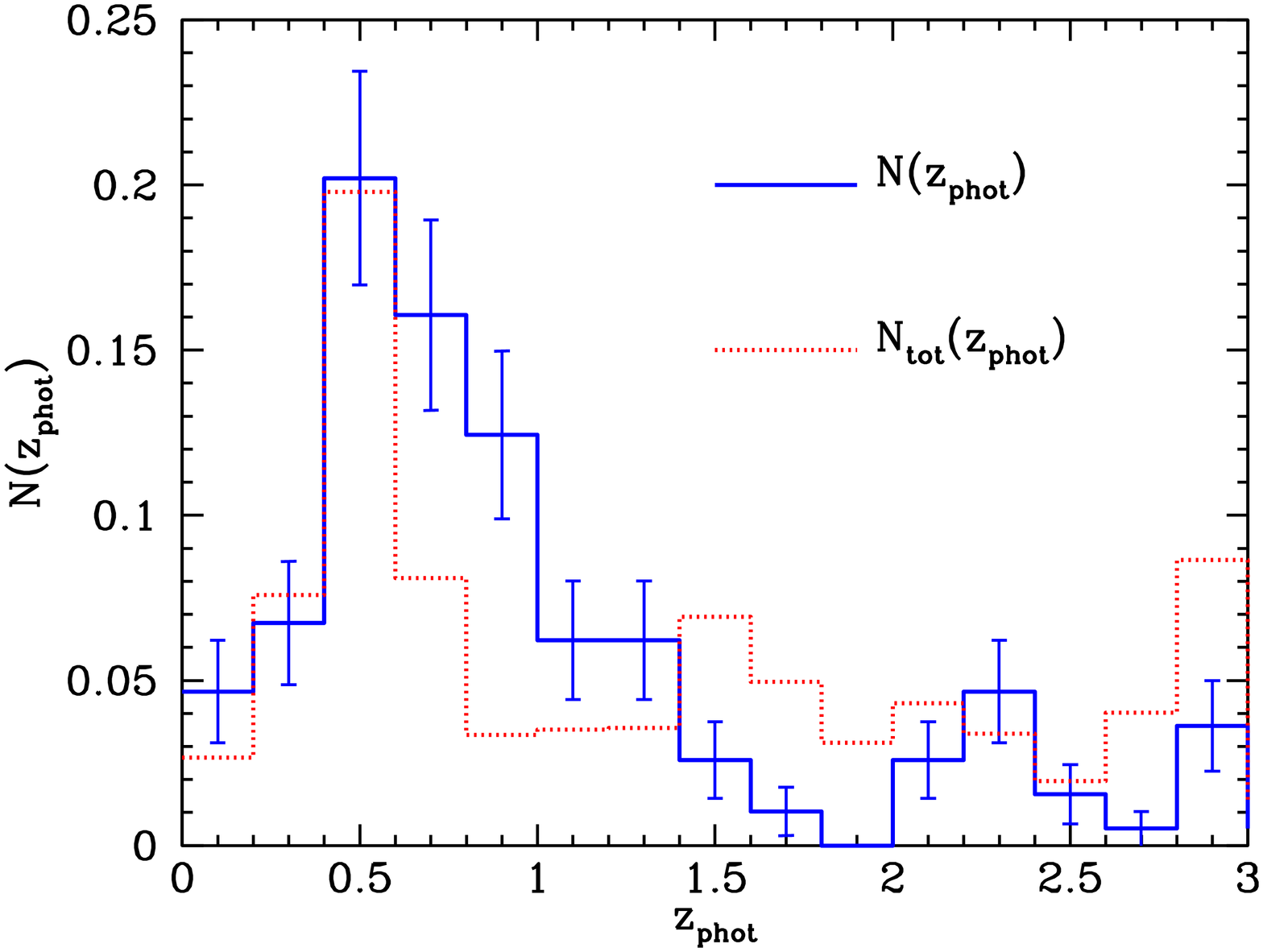}}
  \centerline{\includegraphics[width=6cm,angle=-90]{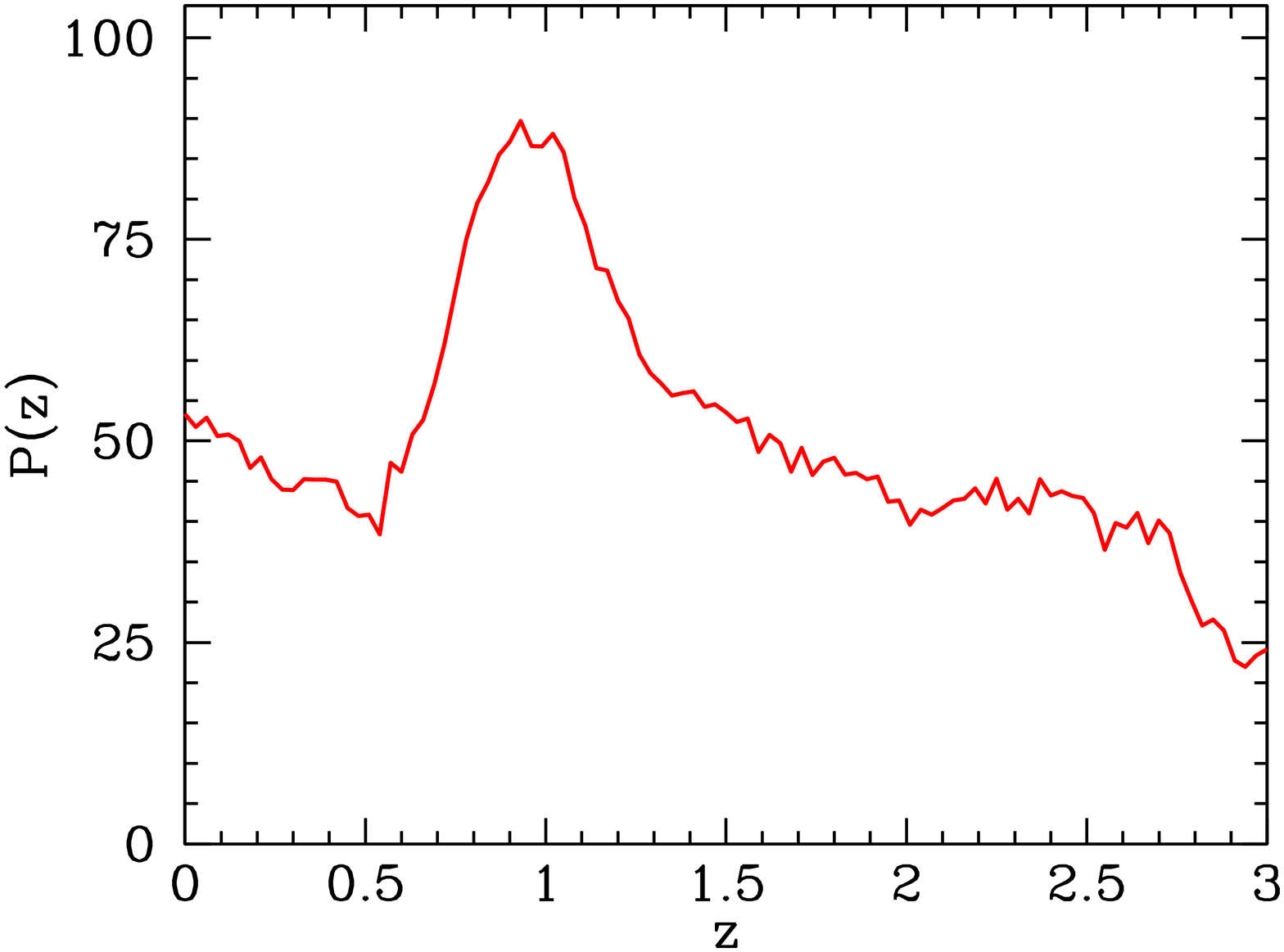}}
  \caption[]{ Photometric redshift determination around the position
    of an X-ray cluster candidate (see text). Top: the photometric
    redshift distribution of the region where the overdensity has been
    detected (solid line), compared to the reference redshift
    distribution, obtained from the total catalogue (dotted line). The
    redshift distributions have been normalised to the respective
    number of objects.  To select the significant overdensities, we
    plot the Poissonian error bars of the small region $N(z_{\rm
    phot})$. Bottom: combined photometric redshift probability as a
    function of redshift of galaxies selected within an aperture of
    radius $1 \arcmin$ whose position within the field maximises the
    probability of the redshift ``peak''. Individual galaxy
    probabilities are determined from the chi-squared statistic
    returned from the photometric redshift template fitting procedure.
    VLT/FORS2 observations showed that this X-ray
    source indeed corresponds to a structure at $z \sim 0.9-1$. See
    Fig. \ref{amas-z1}
\label{photo-z}}
\end{figure}

\begin{figure}[ht]
  \centerline{\includegraphics[width=10cm,angle=0]{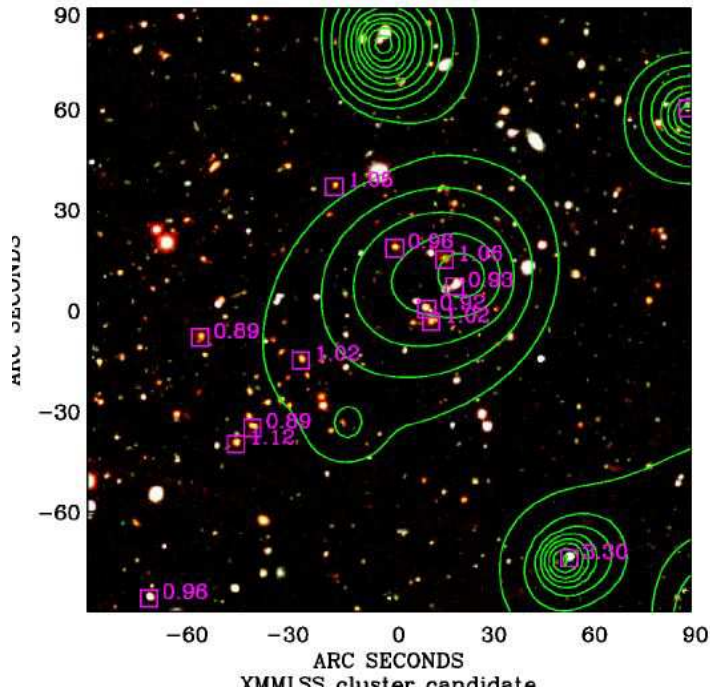}}
  \caption[]{The most distant X-ray complex identified so far, showing
    a clear concentration of galaxies in the $0.9<z<1$ range (Green
    X-ray contours overlaid on a CFH12K VRI composite). Bottom right
    is the most distant X-ray QSO currently measured in the
    XMM-LSS at a redshift $z = 3.3$ (see Fig. \ref{qso}); its
    isophotes are distorted by the immediate vicinity of an X-ray
    bright emission line galaxy located at $z =
    0.054$.\label{amas-z1}}
\end{figure}

The depths of both the XMM-LSS and of the CFHTLS and CTIO data,
have been tuned to sample most of the cluster population out to a
redshift of 1 (Sec. 2.1 and Tab. \ref{flwp}). Beyond this, massive
clusters can still be detected in the X-ray band (Valtchanov et al
2001). A cluster located at $z \sim 2$ and having a temperature of
7 keV will show up with an apparent temperature of about 2.5 keV,
which falls in the most sensitive part of the XMM response. Hence,
for such a cluster, the X-ray K-correction is only $\sim0.7$
\citep{jon}. In contrast, $z>1$ clusters do not appear as
significant overdensities of galaxies in the optical band.
Increasing the depth of the optical survey would not significantly
improve the situation, as most of the galaxy light is shifted into
the infrared band. We have therefore adopted the following
approach and definitions:

1) A $z \leq 1$ cluster candidate is defined as an extended X-ray
source corresponding to a significant visual excess of galaxies in
the optical wavebands. The multi-colour BVRIz' information is used
to enhance the visual contrast of the cluster galaxy density with
respect to the background population and to construct optical
cluster catalogues following different methods : the Red Cluster
Sequence (\citet{glad}, \citet{and}), clustering analysis in
redshift slices \citep{ada}, and matched filter \citep{lob}.These
catalogues are currently used as ancillary information and
cross-correlated with the X-ray extended source catalogue.
Further, the optical data allow us to assign photometric redshifts
to the X-ray cluster candidates in an independent way. First, we
apply the public code Hyperz \citep{bol00} to the entire BVRI
galaxy catalogue. For each galaxy, we derive the photometric
redshift, corresponding to the minimum $\chi^2$ computed comparing
the observed photometry to the fluxes expected from a set of
reference templates (GISSEL98, \citet{bru}). Second, we search for
the cluster signature by placing a number of apertures in the
field of the X-ray detection in order to determine the maximum
overdensity in photometric redshift space relative to the
``field'' distribution   . This approach does not assume that the
distribution of cluster galaxy members apparent in the optical
data is centred on the X-ray detection and it makes no prior
assumption of the apparent optical extent of the candidate
cluster. The redshift bin in $N(z_{\rm phot})$ has been set
similar to the precision expected from photometric redshifts,
given the set of filters and their depths (typically $\Delta z =
0.2$). Finally, from the region with the largest overdensity
detected, we selected the candidate cluster members in different
classes of confidence, considering their $z_{\rm phot}$, their
error bars and probability functions. The method is  illustrated
by Fig.\ref{photo-z} and \ref{amas-z1}.  This procedure enables
the pre-selection of the cluster candidates into {\em NEAR} and
{\em MID} distance classes, corresponding to the $0<z<0.5$ and
$0.5<z<1$ ranges respectively, and provides a useful tool for
ascribing targets to the various telescopes available for the
spectroscopic follow-up.

2) A small fraction of the extended sources ($\leq 1$ per
pointing) do not present any significant optical counterpart --
occasionally, one or two galaxies visible in the I or z' band,
sometimes nothing at all. These are typically interesting $z>1$
cluster candidates and are classified into the third category,
{\em DISTANT}, to be imaged in the NIR bands. If confirmed, they
are the subject of dedicated spectroscopic follow-up programmes.

\subsection{AGNs and QSO}

Active galactic nuclei constitute by far, the dominant population
of X-ray sources at the XMM-LSS sensitivity. Given the XMM PSF,
and consequently, the positional accuracy, there is some ambiguity
in the identification of pointlike sources, as can be appreciated
in Fig. \ref{cluster1} (note the sources having two or three
possible optical counterparts, or for which the optical ID is
below the limit of the optical survey). This is a well known
draw-back, thoroughly studied in deep surveys; but here, the
situation is simpler since the XMM-LSS flux limit is well above
the confusion limit.

\section{First spectroscopic campaigns}
\subsection{Strategy}

The first spectroscopic observations dedicated to the
identification and redshift measurement of the XMM-LSS galaxy
clusters took place during the fall of 2002. Three nights were
allocated at ESO on the VLT/FORS2 instrument and 2 nights at Las
Campanas with the Magellan/LDSS2 instrument. We summarise here the
most important outcomes. Galaxy cluster candidates were
pre-selected as described above, comprehensive information (X-ray,
optical, photometric redshift) being gathered into an HTML
database, allowing a rapid overview of the cluster properties.
From this, we selected a number of clusters spanning different
optical morphology and flux ranges, in order to optimise the
allocated spectroscopic time, and to give a representative
overview of the XMM-LSS cluster population. For the first
spectroscopic campaign, it was decided to conservatively obtain
two masks per cluster in order to sample well the cluster galaxies
in the central region, and ensure confident redshift
determination. Thanks to the high throughput of the FORS2
instrument in the [0.8-1] $\mu$m range, all {\em MID} galaxy
cluster candidates were directed to the VLT sample. The {\em NEAR}
ones were assigned to the Magellan sample, and also as backup
sources for VLT, in case of possible  poor weather. In order to
investigate and quantify the most efficient method for the future,
galaxies to be measured were selected using different approaches:
simple magnitude cut, colour selection or photometric redshift
criteria.

No dedicated spectroscopic runs were planned in 2002 for the QSO
and AGN population. However, many pointlike X-ray sources
surrounding the selected clusters and present in the corresponding
instrumental field of view were included in the spectroscopic
masks. For each $7\arcmin\times 7\arcmin$ field centred on a
galaxy cluster candidate, we first cross-correlated the X-ray
point-like source catalogues with the optical ones obtained from
the CFH12K data. We then searched for optical counterparts within
a radius of $5\arcsec$ from each X-ray point-like source and
produced catalogues of associations. Using the latter, we overlaid
on the optical images the positions of the X-ray sources and their
associated optical counterparts, if any. Then, we visually checked
on the optical images each X-ray position (no more than ten per
field) associated with 0, 1 or more optical counterparts.
Depending on the ``quality'' of the X/optical associations, we
sorted the AGN/QSO candidates into different categories
characterising the priority for spectroscopic observations of the
optical counterparts. There were three main classes: (1)
unambiguous cases where a unique and relatively bright optical
source lies within a radius of $5\arcsec$; (2) ambiguous or
doubtful cases when there were several possible optical
counterparts and/or when the optical counterpart(s) was (were)
rather faint; (3) rejected cases when there was no, or a much too
faint, optical counterpart. The fraction represented by each class
was $\sim$ 35, 25 and 40\% respectively. A catalogue containing
the first two classes of AGN/QSO candidates was considered for
spectroscopic observations of each field. Given the constraints
imposed by the cluster spectroscopy (especially, optimal sampling
of the cluster cores) and the zones of avoidance within the FORS2
field of view, an average of 2-3 pointlike sources were measured
per cluster field. This corresponds to about 30-45 sources per
\dd.

\subsection{Overview of the results}

\begin{figure}[ht]
  \centerline{\includegraphics[width=10cm,angle=0]{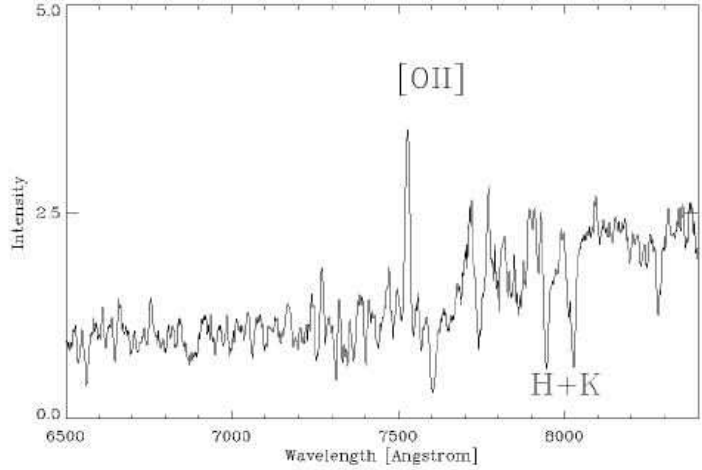}}
  \caption[]{One of the most distant cluster galaxy measured, at a
    redshift of $z = 1.02$ (see Fig. \ref{amas-z1}). This good quality
    spectrum was obtained in 1.5h with the R600RI+19 grism on the
    VLT/FORS2 instrument. The object has magnitudes of V=24.19,
    R=23.34, I=22.21 and displays characteristics of an elliptical
    galaxy having a component of young stars (E+A
    type).\label{galaxy}}
\end{figure}

Weather and working conditions at the VLT were optimal. In three
nights, 12 cluster fields (5 {\em MID}, 7 {\em NEAR} -- $z>0.35$)
were observed, containing on average about 30 slits per mask,
yielding some 700 spectra. In addition, 2 new nearby compact
groups were also observed (with 7 and 8 galaxies per group),
prepared as a backup programme. The overall strategy proved
successful. An example  spectrum is shown in Fig. \ref{galaxy}.
Encouragingly, an estimate of the velocity dispersion (16 cluster
galaxy spectra) of a newly discovered 0.85 redshift cluster can be
achieved in 2 hours, supporting the follow-up of large numbers of
distant clusters with relatively little observing time. Seven {\em
NEAR} clusters were observed at Magellan, where one night out of
the allocated two suffered from poor meteorological conditions.
With an average number of slits of 15 per masks, some 200 spectra
were obtained.

\begin{figure}[ht]
  \centerline{\includegraphics[width=9cm,angle=0]{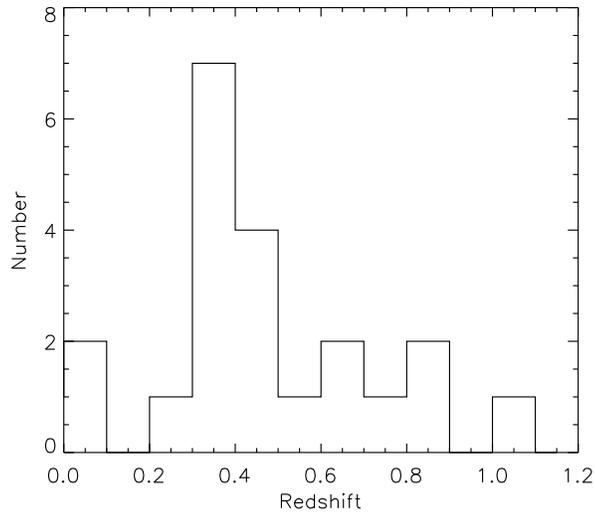}}
  \caption[]{Redshift distribution of the 21 spectroscopically
    confirmed clusters and compact groups during the 2002 Magellan/LDSS and VLT/FORS2
    runs. \label{n_z}}
\end{figure}

\begin{figure}[ht]
  \centerline{\includegraphics[width=10cm,angle=0]{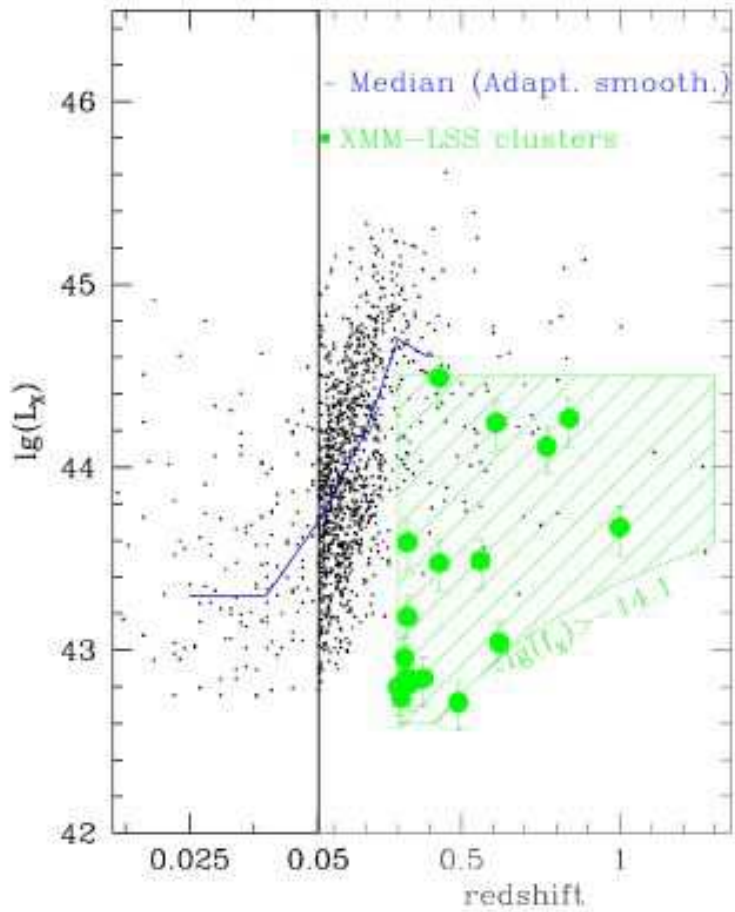}}
  \vspace{5cm}
  \caption[]{The locus of the XMM-LSS clusters. Clusters observed
  during the first 2002 spectroscopic campaign are indicated as
  large filled green circles, while dots indicate all clusters in
  the literature having X-ray observations  (BAX: http://bax.ast.obs-mip.fr/).
  The curved lower edge
of the shaded region indicates the expected XMM-LSS  ultimate
cluster completeness
  limit ($\sim 8 \times 10^{-15} erg/s/cm^{2}$ in [0.5-2] keV),
  whereas the upper limit is merely to guide the eye (maximum
  luminosity at which at least one cluster is expected within the
  full 64 \dd\ area, according to $\Lambda$CDM model). Although
  the 2002 exploratory campaign was not intended to be flux limited, the figure
  well enlightens  the tendency of the
  XMM-LSS to identify significantly lower luminosity (i.e. lower mass) systems
  than the previous cluster surveys. Luminosity is expressed in the
[0.1-2.4] keV band\label{Lxz}}
\end{figure}

The photometric cluster distance estimates proved useful and
reliable indicators for the cluster classification (cf Fig.
\ref{photo-z}, \ref{amas-z1}). Current results on the measured
redshift distribution are shown on Fig.\ref{n_z}. Given the
exploratory selection procedure adopted for this first
identification campaign, this cannot be ascribed any cosmological
interpretation yet.  These measurements, however,  enabled us to
compute the cluster luminosities and, subsequently, to confirm our
sensitivity predictions \citep{ref} and to demonstrate that we are
sampling a new population of low luminosity (i.e. low mass)
objects. Fig. \ref{Lxz} displays the luminosity distribution of
our current confirmed cluster sample as a function of redshift,
compared to the previously known set of X-ray clusters: out to
0.5, we identified a bunch of very faint clusters, thus
significantly extending the sample  that was inventoried in the
very local universe so far ($z<0.1$). Between $0.5<z<1$, we are
similarly discovering the high redshift counterparts of the known
$z<0.5$ cluster population. Further statistical analysis of this
population will be especially relevant to better constrain the
evolution of cluster scaling laws as to the gas properties, since
groups are more sensitive than clusters to the interactions with
the individual galaxies. From the LSS point of view, low mass
groups will provide a precious link between structure underlined
by clusters and those by galaxies, thus enlightening the bias
controversy. A comprehensive X-ray/optical study of the current
{\em NEAR} and {\em MID} samples is presented by \citet{wil} and
\citet{val03} respectively. The optical properties of the cluster
galaxies (luminosity function and colour distribution) are
discussed by \citet{and}.

\begin{figure}[ht]
  \centerline{\includegraphics[width=10cm,angle=0]{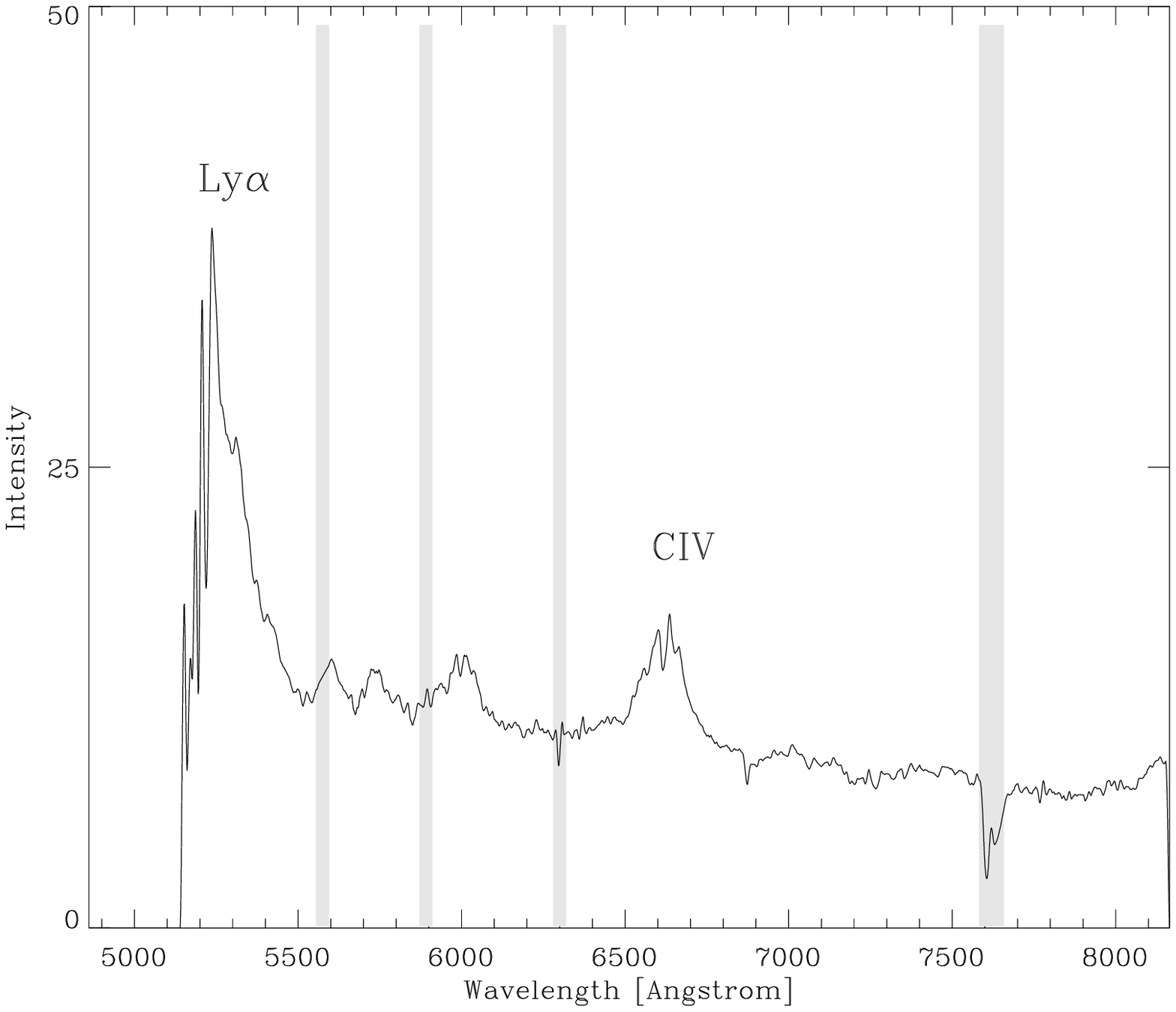}}
  \caption[]{Most distant X-ray AGN detected so far, at a redshift of
    $z = 3.3$ (see Fig. \ref{amas-z1}); grey bands indicate the
    position of strong atmospheric lines.\label{qso}}
\end{figure}

The optical spectra of the measured X-ray AGNs showed a plethora
of properties. The most distant X-ray source measured so far has a
redshift of 3.3 (Fig. \ref{qso}). The AGN sample is currently
under analysis.

\section{First NIR imaging campaign}

\begin{figure}[ht]
  \centerline{\includegraphics[width=10cm,angle=0]{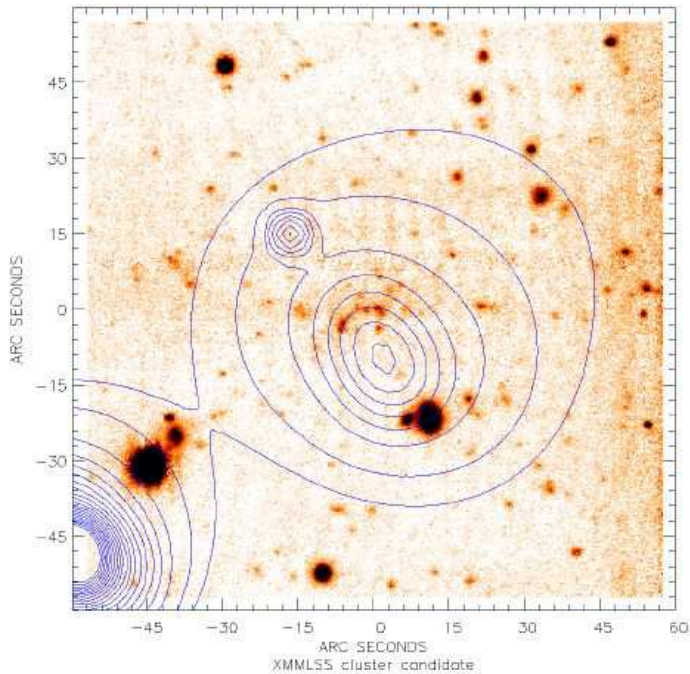}}
  \caption[]{ Extended X-ray source with a very faint optical
    counterpart in the I-band. The NIR image of the field is shown in
    Fig. \ref{dist_k}. \label{dist_i}}
\end{figure}

\begin{figure}[ht]
  \centerline{\includegraphics[width=8cm,angle=0]{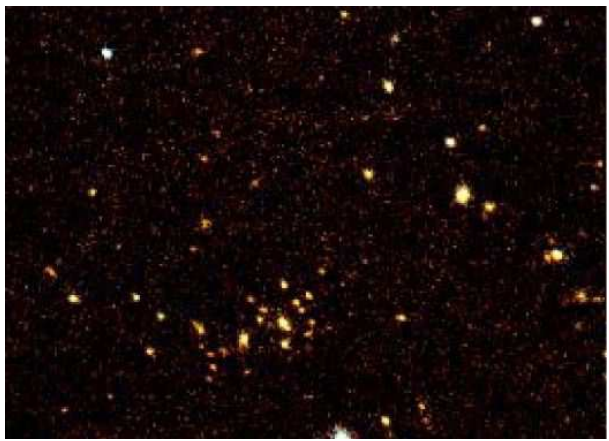}}
  \caption[]{J-K colour image of the {\em DISTANT} cluster candidate
    flagged in Fig. \ref{dist_i}. This presents a significant
    overdensity of red objects. The brightest cluster member has
    Ks$\sim 17.5\pm 0.3$, and the other objects within $20\arcsec$ are
    generally fainter than Ks $ \sim 18.5$. These properties are
    consistent with a cluster at $z=1$ or above. \label{dist_k}}
\end{figure}

In November 2002, we carried out an initial, exploratory observing
run with SOFI on the NTT, in order to search for {\em DISTANT}
clusters selected from candidates in the first set of available
XMM data. The details will be presented  by \citet{and05}. Here we
note that by selecting faint, extended X-ray sources associated
with either blank fields in the CFHT data, or with fields showing
a hint of clustering at faint ($I>22$) magnitude levels in the
same data, we were successful at reliably detecting such clusters.
Moreover, the X-ray fluxes of these objects are such that if they
were placed at higher redshifts they would still be detectable in
the XMM-LSS exposures. Thus the survey will enable us to map the
cluster distribution to redshifts well above $z = 1$. An example
is shown in Fig. \ref{dist_i} and \ref{dist_k}.

\section{Conclusion and prospects}
\subsection{The newly discovered clusters}

Following the completion of the AO-1 period, the main outcome of
the XMM-LSS project can be summarised as follows : XMM
observations of only 10 ks, coupled with a similar time spent on
imaging at CFHT or CTIO, can detect a significant fraction of the
cluster population out to $z\sim 1$ (a flux of $0.8\times10^{-14}$
erg~cm$^{-2}$s$^{-1}$ corresponds to a cluster mass of $6.7\times
10^{13}$ M$_{\odot}$ [T = 2.7 keV] at $z=1$, assuming typical
scaling laws). Moreover, adding two hours of FORS2 spectroscopic
time provides a reliable estimate of cluster velocity dispersion
at $z = 0.85$. This represents a substantial increase in
efficiency compared to former high-z cluster searches.

For the first time we are detecting the numerous population of low
luminosity (mass) objects out to $z\sim 0.5$; this will allow a
dense mapping of the matter distribution. So far (within $\sim$ 4
\dd, including the GT area), no massive object has been detected
and, following a detailed inspection of the optical images, no
giant arcs have been found in the identified clusters. These
results are consistent with the Press-Schechter formalism (folded
with the current favoured cosmological model) predicting 2.8 and
0.0001 cluster per \dd\ between $0<z<1$ for clusters more massive
than $10^{14} M_\odot$ and $10^{15} M_\odot$ respectively, if we
assume that  only massive clusters are potential strong lenses. Of
course, an additional necessary condition is the steepness of the
mass profile.

Immediate foreseen improvements are the following: in the next
version of the X-ray pipeline, the source flux detection limit
will be significantly lowered.  This requires further work on the
background estimation. In parallel, extensive simulations are
being performed to improve the characterisation (extent and flux)
of the faintest sources as well as to determine the survey
selection function. This is complicated by the necessity of taking
proper account of the Poisson nature of the data and the fact that
the signal to be analysed comes from 3 different detectors.

\subsection{Cluster identification refinements}

In a second step, we shall systematically investigate overlaps and
differences between the optical cluster catalogues and the X-ray
extended source catalogue. There are two main reasons why
variations should be observed: (i) faint groups or distant
clusters may not be unambiguously detected as extended sources
because of their low flux or, alternatively, because they host a
cooling flow, which makes them appear unresolved. Currently, we
are confident that any regular object having a typical core radius
of the order of $125~ h^{-1}_{100}$ kpc and producing a net number
of pn counts of $ \geq 70 $ is detected by the pipeline as an
extended source \citep{val01}; (ii) there are intrinsic
differences between the X-ray and optical catalogues since the
methods used to construct them rely on specific assumptions as to
galaxy colour and evolution, or on the properties of the IGM at
high redshift. Understanding the discrepancies will not only
increase the efficiency of our cluster finding procedure, but also
shed light on the much debated topic of cluster formation and
evolution. This will also contribute to the improvement of our
photometric redshift determination procedure. Finally, output from
the weak lensing analysis will reveal large mass concentrations,
optimally within the $0.1<z<0.5$ range. Again, the comparison
between the X-ray and optical catalogues will be most instructive.
Given the current understanding of structure formation, it is
difficult to devise a physical process that would prevent gas
being trapped or heated within large concentrations of dark
matter. However, the surveyed volume is ideally suited to
systematically search for ``dark clump''.

\subsection{Active galactic Nuclei}

Due to the primary goal of the spectroscopic observations
performed so far (confirmation and redshift determination of X-ray
selected galaxy clusters), few AGN/QSO candidates could   be
observed. For this reason, a spectroscopic survey of a unique
sample of X-ray selected AGN/QSOs within a large contiguous area
of some 10 \dd, typically complete down to F$_{[2-10] keV}
\sim10^{-14}$~erg~cm$^{-2}$~s$^{-1}$, F$_{[0.5-2] keV} \sim
5\times10^{-15}$~erg~cm$^{-2}$~s$^{-1}$ and I$_{AB}$ = 22.5 is
foreseen. Because of the large surface density of AGN/QSO
candidates detected in the XMM-LSS field (typically
$>200$~deg$^{-2}$ with $0< z < 1$), compared to previous surveys,
it should be more sensitive to contrasts between voids and peaks
than previous surveys, such as 2dF. This will enable us to study
with high precision the correlation function of these objects over
scales in the [2-400] Mpc range, to probe the environmental
influence on various type of AGN, and to compare these
correlations with those for normal galaxies. The distinct redshift
distributions of X-ray selected type I and type II AGN can be
accurately determined, and this will provide interesting
constraints on models of black hole formation compared with models
of star formation. The proposed studies will provide a comparison
of the clustering properties, as a function of redshift, of X-ray,
optical (2dF) or radio selected AGN/QSO.

\subsection{Survey products}

A dedicated {\tt mysql} database with a Java front end interface
is available via a site at IASF
Milano\footnote{http://cosmos.mi.iasf.cnr.it/$\sim$lssadmin/Website/LSS/}
with a mirror at ESO Santiago. This includes the X-ray source
catalogue, and complete catalogues of surveys performed by the LSS
Consortium in other wavebands, together with selected subsets from
the surveys performed by other Consortia made available under
agreement. It  also gives access to a selection of data products.
The database, currently accessible and used internally by the
Consortium, will be gradually opened to the public. The first
public release, based on the AO-1 pointings (X-ray source lists
and available identifications) is foreseen by the end of 2004.

\acknowledgments{XMM is an ESA science mission with instruments
and contributions directly funded by ESA Member States and NASA.
SDS was supported by a post-doctoral position from the Centre
National d'Etudes Spatiales. MP and IV are grateful to the
ESO/Santiago Office for Science, for a 2 week stay in October
2002, where the analysis of the VLT data presented here was
initiated; they would like to thank G. Marconi, for his support at
the VLT. Thanks also to G. Boese and R. Gruber at MPE, for
enlightening discussions about the RASS PSF}

\end{document}